\documentclass[%
reprint,
groupedaddress,
 amssymb,
 aps,
prc,
amsmath,
]{revtex4-1}

\usepackage{graphicx}
\usepackage{dcolumn}
\usepackage{bm}
\newcommand{\braket}[1]{\left\langle #1 \right\rangle}

\newcommand{\ket}[1]{\left| #1 \right \rangle}

\usepackage{bm, color}
\usepackage{lipsum}

\definecolor{ryangreen}{rgb}{0.20,0.8,0.0}

\definecolor{red}{rgb}{0.81,0.13,0.16}

\usepackage{hyperref}
\hypersetup{
    colorlinks=true,
    linkcolor=blue,
    filecolor=magenta,      
    urlcolor=blue,
    citecolor=blue,
    }

\begin{document}

\preprint{APS/123-QED}

\title{Ab initio study of the neutron and Fermi polarons on the lattice}

\author{Ryan Curry}
\author{Jasmine Kozar}
\author{Alexandros Gezerlis}
\affiliation{Department of Physics, University of Guelph, Guelph, Ontario N1G 2W1, Canada}

\date{\today}

\begin{abstract}
      We have used the auxiliary-field quantum Monte Carlo (AFQMC) many-body approach on the lattice to study the equation of state for a fermionic impurity interacting with a background sea of spin-polarized fermions. 
      The impurity, or polaron, is an interesting system in both cold atomic and nuclear physics. Our approach is general, and we are able to straightforwardly study the polaron across these regimes. We first study the Fermi polaron at unitarity and for a wide range of scattering lengths, comparing against previous theoretical and experimental studies. We then explore the neutron polaron which has been shown to be an important constraint for nuclear physics.
      We have also employed the recently developed parametric matrix model to emulate AFQMC solutions to the two-body problem on the lattice, to accelerate the tuning of our lattice Hamiltonian parameters directly to two-body energies in a periodic box, following L\"{u}scher's formula.
      Our lattice quantum Monte Carlo results for the polaron in both a cold atomic and nuclear physics context can serve as stringent benchmarks for future theoretical and experimental research. 
\end{abstract}

\maketitle
\section{Introduction}
The polaron, a quasiparticle first proposed by Landau and Pekar to describe conduction electrons amidst an ionic lattice \cite{Landau:1933, Landau:1948ijj} and more recently come to describe a single impurity interacting with a background sea of opposite-spin fermions \cite{Schirotzek:2009aa}, is now a widely studied problem in many areas of physics such as ultracold atomic gases, condensed matter physics, and nuclear physics \cite{Massignan:2014qug, Franchini:2021aa, Tajima:2023hsc}. It also presents an excellent opportunity to test quantum Monte Carlo (QMC) methods, as the extreme population imbalance leads to a severe fermion sign problem (see Fig.~\ref{fig:constrained}). 

In nuclear physics, the neutron polaron has been used to provide a constraint for energy-density functionals in the limit of extreme polarization \cite{Forbes:2013ava, Roggero:2014cma} and the proton polaron could prove useful in the study of neutron stars, where a system consisting of a single proton immersed in a background of neutron matter provides insight on the neutron-rich region of the star's outer core \cite{Kutschera:1992ua, Roggero:2014cma, Tajima:2023gvu}. The physics governing an impurity also likely plays a role in the study of nuclei, which can exhibit exotic behavior such as neutron halo structures and $\alpha$-particle clustering \cite{Zhou:2009sp, Tanaka:2021oll}.

While extremely neutron-rich matter is not accessible with terrestrial experiments, its study can be informed through the use of experiments in ultracold atomic gases. Due to the large negative scattering length of the s-wave neutron-neutron interaction, there is a natural comparison with studies of cold atoms which are able to tune the scattering length between particles through the use of Feshbach resonances \cite{Chin:2010crf}. 
This has allowed a number of cold-atomic experiments to probe the Fermi polaron in different contexts \cite{Schirotzek:2009aa, Yan:2019oqn, Fritsche:2021bsr}.

In addition to the many experiments investigating the polaron, it has also been the subject of a number of theoretical studies, uncovering phenomena such as a polaron to molecule transition \cite{Prokofev:2007aa, Peng:2021aa} and a phase separation \cite{Pilati:2007inu}. 
The Fermi polaron has also been studied using many-body techniques including phenomenological variational approaches \cite{Mora:2009gis, Punk:2009aa}, diagrammatic Monte Carlo \cite{Prokofev:2007aa, VanHoucke:2020aa} as well as QMC methods \cite{Lobo:2006aa, Bour:2014bxa, Pessoa:2021aa}. 
While less intensely studied, the neutron and proton polarons have been studied using QMC \cite{Forbes:2013ava, Roggero:2014cma}, the Brueckner-Hartree-Fock approach \cite{Vidana:2021tiw}, and with the many-body T-matrix approach \cite{Tajima:2023gvu}.

QMC methods are currently one of the most powerful approaches for solving the many-body Schr\"{o}dinger equation. 
While initially being cast in coordinate space with purely central interactions \cite{Kalos:1962xxb, Anderson:1975aa}, there are now a number of QMC approaches for nuclear physics, in coordinate space \cite{Carlson:1987zz, Schmidt:1999lik}, on the lattice \cite{Lee:2008fa, Curry:2023sxh, Lee:2025req}, and in Fock space \cite{Roggero:2014lga}. One of QMC's most attractive features for nuclear (and cold-atomic) physics is the ease with which a single Hamiltonian can be used to study diverse systems ranging from the lightest nuclei to systems in the thermodynamic limit such as neutron matter and symmetric nuclear matter \cite{Gezerlis:2009iw, Gandolfi:2014ewa, Lonardoni:2019ypg}.
QMC methods can also straightforwardly handle systems with large spin or mass imbalances \cite{Gezerlis:2010ww, Gezerlis:2009xp, Gandolfi:2024stu}. 

Our approach to investigating the polaron on the lattice involves the use of the AFQMC method, which was first developed for condensed matter systems \cite{Blankenbecler:1981jt, Sugiyama:1986aa, Zhang:1996us} where it has been very successful \cite{Zhang:2003zzk, Shi:2013aa, Qin:2016aa}.
Recently, it has been extended to study cold-atomic systems \cite{Carlson:2011kv, Shi:2015aa} as well as low-density neutrons \cite{Curry:2023sxh}.
There also exist variants of the AFQMC method that have been generalized to handle finite temperature calculations \cite{Magierski:2008wa, Jensen:2019zkr, Jensen:2018opr, Lu:2019nbg, Ren:2023ued}.

To show the generality of our lattice AFQMC approach, this work studies both the neutron polaron and the cold-atomic polaron using a single many-body Hamiltonian. After a brief review of the AFQMC method in Sec.~\ref{sec:afqmc}, we introduce a new method for tuning our lattice interaction in Sec.~\ref{sec:PMM} employing an approach the combines L\"{u}scher's formula for two particles in a finite volume \cite{Luscher:1986pf, Luscher:1990ux} and recently developed emulators \cite{Cook:2024toj, Somasundaram:2024zse}.
This allows us to study the polaron in Sec.~\ref{sec:polaron} in the contexts of nuclear physics and ultracold atoms. 
We study the cold-atomic polaron both at unitarity and across the BCS-BEC crossover, finding excellent agreement with both experiment and previous many-body calculations. Our lattice calculations of the neutron polaron are the first of their kind and provide an updated constraint over a range of Fermi momenta. These calculations also exhibit how AFQMC neatly handles the fermion sign problem and provide guidance for future experimental investigations.

\section{Auxiliary-Field Quantum Monte Carlo} \label{sec:afqmc} 
Our lattice Hamiltonian is written in second quantization and takes the form,
\begin{align} \label{eq:H}
	\hat{H} = \frac{1}{M^3} \sum_{\bm{k}, i,j, \sigma} \hat{c}_{i\sigma}^{\dag} \hat{c}_{j\sigma} \epsilon_{k} e^{i\bm{k}\cdot(\bm{r}_i - \bm{r}_j)} + U\sum_{i} \hat{n}_{i\uparrow}\hat{n}_{i\downarrow},
\end{align}
where $M^3$ is the number of the lattice sites and the $\hat{c}^{\dag}_{i \sigma}$ and $\hat{c}_{j \sigma}$ are creation and annihilation operators for a particle of spin projection $\sigma$ on the lattice sites $i$ and $j$. Since space is discretized on a lattice, there is also a corresponding momentum lattice  where the $N_k^3$ momentum points are given by $\bm{k} \equiv \frac{2\pi}{L}(n_x \hat{\bm{x}} +  n_y \hat{\bm{y}} + n_z \hat{\bm{z}})$ and the number of momentum lattice points is determined by the number of coordinate space lattice points, i.e. $N_k = (M-1)/2$. The kinetic energy dispersion relation $\epsilon_{k}$ is given by,
\begin{align} \label{eq:edisp}
	\epsilon_{k} = \frac{k^2}{2m} [1 + \gamma k^2 \alpha^2], 
\end{align}
where $\alpha = L/M$ is the lattice spacing, and the $\gamma$ term has be introduced to modify the effective range of the interaction. Note that our calculations are carried out in units where $\hbar=1$. Together with $\gamma$, the interaction strength $U$ (which is always negative for our attractive interactions) can be tuned to vary the scattering length of the interaction. The values of these parameters for an interaction describing cold atoms are known \cite{Werner:2012xbg, Carlson:2011kv} and have also been tuned to reproduce the neutron-neutron s-wave scattering length and effective range in Ref.~\cite{Curry:2023sxh}. 
In Sec.~\ref{sec:PMM} we discuss a new approach to tuning these parameters for s-wave nuclear interactions using L\"{u}scher's formula and newly developed emulators. 

An important distinction between AFQMC and other quantum Monte Carlo approaches is that it is cast in a Slater determinant space, as opposed to coordinate or momentum space. A general Slater determinant is written as the antisymmetrized product of single particle orbitals,
\begin{align} \label{eq:SD}
	\ket{\phi} = \hat{\phi}_1^{\dag} \hat{\phi}_2^{\dag} \cdots \hat{\phi}_N^{\dag} \ket{0},
\end{align}
where $N$ is the number of particles and $\hat{\phi}_j^{\dag} = \sum_i \hat{c}_i^{\dag} \phi_{i,j}$ creates a particle in the $j^{\text{th}}$ single particle orbital.  
The full many-body wave function is written as,
\begin{align} \label{eq:Phi}
	\Phi = \begin{pmatrix}
	\phi_{1,1} & \phi_{1,2} & \cdots & \phi_{1,N}\\
	\phi_{2,1} & \phi_{2,2} & \cdots & \phi_{2,N}\\
	\vdots & \vdots & \ddots & \vdots\\
	\phi_{M^3,1} & \phi_{M^3,2} & \cdots & \phi_{M^3,N}
	\end{pmatrix},
\end{align}
where $\Phi$ is a non-square matrix with dimensions $M^3 \times N$. 
The matrix elements of $\Phi$ are the $\phi_{ij}$ coefficients that appear in Eq.~(\ref{eq:SD}), meaning each column of $\Phi$ is a vector describing a single particle orbital. In contrast to other diffusion Monte Carlo approaches \cite{Schmidt:1999lik, Gezerlis:2007fs} in which the Monte Carlo walkers are particle configurations, in AFQMC it is these vectors which are propagated as random walkers.  

AFQMC is one of a number of QMC methods \cite{Carlson:1987zz, Schmidt:1999lik, Lee:2008fa} that involve recasting the many-body Schr\"{o}dinger equation in terms of a diffusion equation in imaginary time and projecting out the ground state wave function, $\ket{\phi_0}$, 
\begin{align} \label{eq:tau-proj}
	\ket{\phi} &= e^{-(\hat{H}-E_T)\tau} \ket{\phi_T}  \nonumber
	\\
        &= \sum_i c_i e^{-(\hat{H}-E_T)\tau} \ket{\phi_i}
	\\
	&= c_0 e^{-(E_0-E_T)\tau} \ket{\phi_0}, \ \ \ \ \ \ \text{lim}\ \  \tau \rightarrow \infty\,. \nonumber
\end{align}
While the exponentials containing excited states, $i>0$, decay as $\tau$ becomes large, the ground state contribution remains finite so long as our trial wave function $\ket{\phi_T}$ is nonorthogonal to the true ground state $\braket{\phi_T | \phi_0} \neq 0$. Choosing a trial energy $E_T$ such that $E_T \approx E_0$ will aid in keeping the wave function normalization under control. For this work we employ a simple free-particle trial wave function found by diagonalizing the one-body kinetic energy part of the Hamiltonian.  

In order to accomplish the imaginary time projection in Eq. (\ref{eq:tau-proj}), we rewrite the full propagation as the product of many infinitesimally small imaginary time steps,
\begin{align}
	\ket{\phi} = \prod_n e^{-(\hat{H}-E_T)\Delta \tau} \ket{\phi_T},
\end{align}
where $n\Delta \tau = \tau$, and the short-time propagator is decomposed using a Trotter approximation giving a single propagation step as, 
\begin{align} \label{eq:prop}
	\ket{\phi}^{(n+1)} &= e^{-(\hat{H} - E_T) \Delta \tau} \ket{\phi}^{(n)} \nonumber
	\\
	\ket{\phi}^{(n+1)} &= e^{-(\hat{K} + \hat{V} - E_T) \Delta \tau} \ket{\phi}^{(n)}
	\\
        &\approx e^{- \frac{\hat{K}\Delta\tau}{2}} e^{- \hat{V}\Delta\tau} e^{- \frac{\hat{K}\Delta\tau}{2}} \ket{\phi}^{(n)} e^{E_T\Delta\tau}, \nonumber 
\end{align}
where $\hat{K}$ and $\hat{V}$ are the kinetic and potential energy operators, and the normalization aspect of $E_T$ is more clear.

The kinetic energy propagation is straightforward, since the propagator for the kinetic energy operator in Eq.~(\ref{eq:H}) is of the form $\text{exp}[\sum_{ij} \hat{c}_i^{\dag} O_{ij} \hat{c}_j]$, where $O_{ij}$ are the matrix elements of a one-body operator, which has been shown to produce another Slater determinant when acting upon a Slater determinant \cite{Hamann:1990aa}.
However the potential energy propagator is not of this form since $\hat{n}_{i \uparrow} \hat{n}_{i \downarrow} = \hat{c}_{i \uparrow}^{\dag} c_{i \uparrow} \hat{c}_{i \downarrow}^{\dag} c_{i \downarrow}$ is quadratic, and so we employ the discrete Hubbard-Stratonovich transformation for attractive interactions to reduce the $\hat{n}$ dependence from quadratic to linear \cite{Hirsch:1983zza}.
The potential energy propagator is rewritten,
\begin{align}
	e^{-\Delta \tau U \hat{n}_{i \uparrow}\hat{n}_{i \downarrow}} &= e^{-\frac{\Delta \tau U}{2}(\hat{n}_{i \uparrow} + \hat{n}_{i \downarrow})} \sum_{x_i = \pm 1} p(x_i) e^{[\lambda x_i (\hat{n}_{i \uparrow} + \hat{n}_{i \downarrow})]} \nonumber
	\\
	&= \sum_{x_i = \pm 1} p(x_i) \prod_{\sigma = \uparrow, \downarrow} e^{-[\frac{\Delta \tau U}{2} - \lambda x_i]c_{i \sigma}^{\dag}c_{i \sigma}}
\end{align}
where $\lambda$ is given by $\cosh \lambda = e^{\Delta \tau |U|/2}$ and the $x_i$ are auxiliary-fields weighted with a probability density function $p(x_i)$. This form of the Hubbard-Stratonovich transformation for attractive interactions leads to an auxiliary-field dependent $p(x_i)$ given by, 
\begin{align}
	p(x_i) = \frac{1}{2} e^{[\frac{\Delta \tau U}{2} - \lambda x_i]}.
\end{align}

It is worth noting this form of the discrete Hubbard-Stratonovich differs from what is used with AFQMC for repulsive interactions. In that case the probability density function is simply given by $1/2$. We have tested moving the auxiliary-field dependence in our $p(x_i)$ to the normalization factor in front of the sum and using the simple $\frac{1}{2}$ weights, however we found this leads to a dramatic increase in the Monte Carlo walker weights (discussed below) which quickly makes the computation impractical.  

To fully propagate our system over all lattice sites, we can write the configuration of auxiliary-fields as $\bm{x}$ and the full imaginary time propagator from Eq.~(\ref{eq:prop}) as,

\begin{align} \label{eq:final-prop}
	e^{-(\hat{H} - E_T) \Delta \tau} \approx e^{\Delta \tau E_T} \sum_{\bm{x}}P(\bm{x}) \hat{B}_{K/2} \hat{B}_V(\bm{x}) \hat{B}_{K/2}
\end{align}
where $P(\bm{x}) = \prod_i p(x_i)$ is a probability density function over the full set of auxiliary fields. The kinetic and potential energy propagators are $M^3 \times M^3$ matrices given by, 
\begin{align}
	\hat{B}_{K/2} = e^{-\Delta \tau \sum_{i,j}\hat{c}_i^{\dag}K_{ij}\hat{c}_j} 
\end{align}
with $K_{ij} = \sum_{\bm{k}} \epsilon_{k} e^{i\bm{k}\cdot(\bm{r}_i - \bm{r}_j)}$, and, 
\begin{align} \label{eq:BV}
	\hat{B}_V(\bm{x}) = \prod_i e^{-[\frac{\Delta \tau U}{2} - \lambda x_i]c_{i}^{\dag}c_{i}}.
\end{align}

We have neglected to include any $\sigma$ subscript in the propagator as the Hubbard-Stratonovich transformation employed treats the spin-up and spin-down particles on equal footing without mixing spins. This means the Slater determinant and propagators can each be split into spin-up and spin-down parts (i.e. $\hat{B}\Phi = \hat{B}^{\uparrow}\Phi^{\uparrow} \otimes \hat{B}^{\downarrow}\Phi^{\downarrow}$).

Writing the imaginary time propagator in the form of Eq.~(\ref{eq:final-prop}) means that the propagation step initially described in Eq.~(\ref{eq:prop}) can be implemented as straightforward matrix multiplications,
\begin{align} \label{eq:mat-mul}
	|\phi^{(n+1)}\rangle = e^{\Delta \tau E_T} \sum_{\bm{x}} P(\bm{x}) \hat{B}_{K/2} \hat{B}_{V}(\bm{x}) \hat{B}_{K/2} |\phi^{(n)}\rangle.  
\end{align}
As mentioned above, to carry out this propagation in a Monte Carlo context the wave function consists of many copies of the initial trial state (walkers) that are all propagated through the Monte Carlo random walking process. The measurement of expectation values, which we discuss in more detail below, is then simply an average over all of the Monte Carlo walkers. To propagate our walkers, we stochastically sample the auxiliary-fields variables $\bm{x}$ and then carry out the matrix multiplication in Eq.~(\ref{eq:mat-mul}). 

As is common in almost all quantum Monte Carlo methods, we also employ an importance-sampling procedure to improve the efficiency of our approach. 
A standard choice is to assign a weight to each walker given by the overlap between its propagated Slater determinant and the trial wave function, 
\begin{align} \label{eq:weights}
w_k = \braket{\phi_T | \phi_k},
\end{align}
where $k$ denotes a specific walker. 
This also leads to a modified sampling for our auxiliary fields. For a given walker we have the probability density function,
\begin{align}
	p(x_i) = \frac{1}{2} \frac{\langle \phi_T | \phi^{(n)}_{k,i} \rangle}{\langle \phi_T |\phi^{(n)}_{k,i-1} \rangle} e^{[\frac{\Delta \tau U}{2} - \lambda x_i]},
\end{align}
which is now influenced by the ratio of the overlaps as we sample through the auxiliary-field variables $x_i$.

The walker weights given by Eq.~(\ref{eq:weights}) can also be used to impose the constrained path approximation \cite{Zhang:1996us} to deal with the fermion sign problem.

\begin{figure}[t]
\centering
    \includegraphics[width=0.49\textwidth]{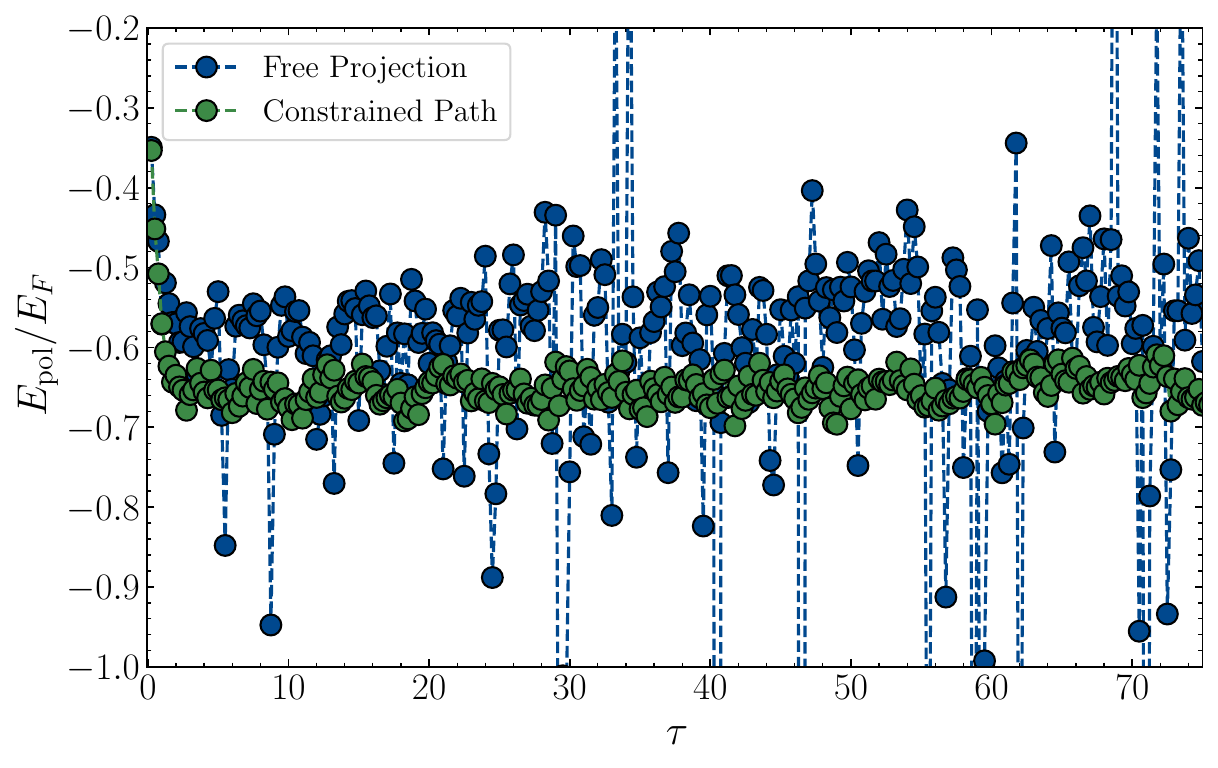}
    \caption{Comparison between constrained path and free projection AFQMC calculations for the polaron at unitary on an $11^3$ site lattice. Dashed lines are included only to guide the eye.}
    \label{fig:constrained}
\end{figure}

The sign problem is notorious in QMC approaches to the many-body problem, and there are a number of different approaches to taming it \cite{Zhang:1996us,Reynolds:1982aa}. 
In coordinate space QMC approaches, the sign problem occurs because the antisymmetric trial wave function introduces a nodal surface $\phi_T=0$, which particles can cross many times during the random walk. 
This leads to a growing amount of statistical noise which drowns out the calculation. 
In a similar fashion, the fermion sign problem occurs for our approach due to a symmetry between the true ground state of a system $\ket{\phi_0}$, and its negative $-\ket{\phi_0}$ \cite{Zhang:1996us}. 
Therefore there exists some nodal surface defined by $\braket{\phi_0| \phi} = 0$, and our Slater determinant walkers can in principle cross this surface. 
The issue arises because once a walker reaches the nodal surface it should in principle no longer contribute to the Monte Carlo estimate, however due to the Monte Carlo sampling the walkers continue to propagate (along either signed side of the nodal surface) and eventually the signal-to-noise ratio of the calculation grows too large.
To avoid this issue, Ref.~\cite{Zhang:1996us} introduced the constrained path approximation which imposes that at each step in the imaginary time propagation, the walkers must maintain a positive overlap with the trial wave function,
\begin{align}
	\langle \phi_T | \phi_k^{(n)} \rangle > 0,
\end{align}
and replacing our importance function in Eq.~(\ref{eq:weights}) with $\text{max}[\braket{\phi_T | \phi_k}, 0]$, which yields an approximate solution for the ground-state wave function by ensuring the distribution of walkers vanishes smoothly at the nodal surface, without crossing it. 
To illustrate the effect of the constrained path approximation, we show in Fig.~\ref{fig:constrained} the results of both a constrained path AFQMC calculation as well as a free projection (no constraint) calculation \cite{Zhang:2003zzk} for the unitary polaron on an $11^3$ lattice. While both calculations agree for very small imaginary times, the signal of the Monte Carlo estimate is lost to noise very quickly as the sign problem accumulates in the free projection calculation. 

To compute expectation values in the constrained path approximation, we consider the mixed estimate, 
\begin{align} \label{eq:mixed}
O_{\text{mixed}} = \frac{\langle \phi_T| \hat{O} | \phi_0 \rangle}{ \langle \phi_T | \phi_0 \rangle} &\equiv \frac{\sum_k w_k \frac{\langle \phi_T | \hat{H} | \phi_k \rangle}{\langle \phi_T | \phi_k \rangle}}{ \sum_k w_k },
\end{align}
which is evaluated with a Monte Carlo sum over all walkers, $k$, and the ground state wave function is estimated by our importance sampled and imaginary time propagated Slater determinants.
For the energy expectation value the mixed estimate in Eq.~(\ref{eq:mixed}) is equal to the pure estimate $\braket{\phi_0 | H | \phi_0}$, however approximations such as extrapolated estimates or back-propagation \cite{Zhang:1996us} are required for operators that do not commute with the Hamiltonian.
The kinetic energy terms inside the sums in Eq.~(\ref{eq:mixed}) for the ground-state energy can be calculated with the one-body equal-time Green's function following,
\begin{align} \label{eq:Gij}
	\langle c_{j \sigma}^{\dagger} c_{i \sigma} \rangle &= \frac{\langle \phi_T| c_{j \sigma}^{\dagger} c_{i \sigma}| \phi_k \rangle}{ \langle \phi_T | \phi_k \rangle}
	\\
	& = [\Phi^{\sigma} [ (\Phi_T^{\sigma})^{\dagger} \Phi^{\sigma}]^{-1} (\Phi_T^{\sigma})^{\dagger}]_{ij},
\end{align}
where $\Phi^{\sigma}$ is spin-decomposed portion of the full wave function from Eq.~(\ref{eq:Phi}) as discussed after Eq.~(\ref{eq:BV}). The potential energy term in our Hamiltonian does not naturally appear in the form required for an application of Eq.~(\ref{eq:Gij}), however it can be represented in an appropriate form through an application of Wick's theorem \cite{Assaad:2002aa, Nguyen:2014nza}.

Now that we have reviewed the details of the AFQMC algorithm, we turn to a discussion of tuning our lattice interaction parameters to study different physical systems.

\section{Emulators for Tuning Lattice Interactions} \label{sec:PMM}
Previous lattice QMC investigations into cold-atomic systems \cite{Carlson:2011kv, Jensen:2019zkr,Curry:2023sxh} used the formalism developed in Refs.~\cite{Werner:2010aa, Werner:2012xbg} to tune the dispersion relation in Eq. (\ref{eq:H}) to have zero effective range. In Ref.~\cite{Curry:2023sxh} we employed this formalism to additionally reproduce the scattering length and effective range parameters of the neutron-neutron s-wave interaction. In this work, we instead opt to tune our lattice interaction via a direct energy comparison with L\"{u}scher's formula \cite{Luscher:1986pf, Luscher:1990ux, Beane:2003da} which relates scattering phase shifts in the continuum and inside a finite box. For two particles in a periodic box of length $L$, L\"{u}scher's formula takes the effective-range expansion,
\begin{align} \label{eq:ere}
	p \cot \delta = -\frac{1}{a} + \frac{1}{2} r_e p^2 + \cdots ,
\end{align}
where $p^2 = mE$, $\delta$ is the s-wave phase shift, and $a$ and $r_e$ are the scattering length and effective range, and re-expresses it as,
\begin{align} \label{eq:lusher}
	p \cot \delta(p) = \frac{1}{\pi L} S\biggl[ \biggl( \frac{Lp}{2\pi} \biggl)^2 \biggl],
\end{align}
where $S(\eta) = \sum_{\bm{j}}^{\Lambda}\frac{1}{|\bm{j}|^2 - \eta} - 4\pi \Lambda$ is a sum that can be evaluated numerically \cite{Klos:2016fdb}. By solving for the $E=p^2/m$ that simultaneously satisfies both Eq. (\ref{eq:ere}) and Eq. (\ref{eq:lusher}) one finds the finite-box ground-state energy for a given scattering length and effective range as a function of $L$. 

The standard approach would be to compute the ground-state energy of the two-body system using AFQMC on several small lattice sizes, tuning the lattice interaction parameters $\gamma$ and $U$ until our results exactly match with the energies found from solving L\"{u}scher's formula, ensuring we are describing the correct two-body physics for a given scattering length and effective range. 
This is accomplished through a least-squares minimization procedure, with a cost function of the form $\sum_{i} (E_{\text{AFQMC}}^i - E_{\text{L\"{u}scher}}^i)^2$, where $i$ iterates over the $7^3$, $9^3$, and $11^3$ lattices, since smaller lattice sizes are known to diverge from L\"{u}scher's formula \cite{Beane:2003da, Curry:2023sxh}.  
Unfortunately, carrying out the least-squares minimization procedure with the full AFQMC calculations for several lattice sizes proves to be quite costly, and the entire process must be repeated each time the interaction is changed. As a potential path to circumvent this slowdown, we can turn to the burgeoning field of emulators for nuclear physics.

The concept behind emulators is to develop an algorithm that reproduces the results of a high-fidelity and often expensive many-body calculation, for a fraction of the computational cost. While there have been a number of emulators developed for nuclear many-body calculations \cite{Frame:2017fah,Ekstrom:2019lss,Konig:2019adq,Melendez:2021lyq,
Wesolowski:2021cni,Djarv:2021hcj, Franzke:2023hdh, Belley:2023lec, Jiang:2022oba, Jiang:2022tzf, Duguet:2023wuh, Somasundaram:2024zse} we have implemented the parametric matrix model (PMM) which was recently proposed \cite{Cook:2024toj} and has already been successfully applied to a number of many-body nuclear physics calculations \cite{Somasundaram:2024zse, Reed:2024urq, Somasundaram:2024ykk, Armstrong:2025tza, Curry:2025pna}. One of the main benefits of the PMM is that it is a data-driven approach, and so does not require any detailed knowledge of the many-body wave function. This makes the PMM particularly attractive for diffusion-based quantum Monte Carlo methods, where there is no direct access to the propagated wave function.

\begin{figure}[t]
\centering
    \includegraphics[width=0.49\textwidth]{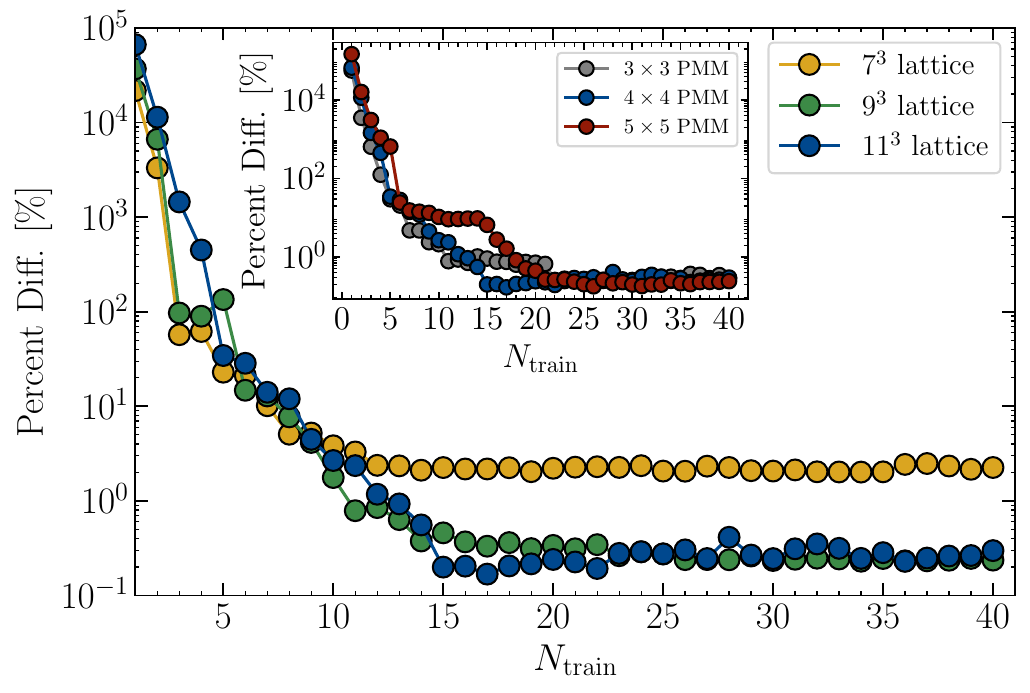}
    \caption{Average percent error of 4$\times$4 PMMs as a function of the number of training points given to the model. Each PMM was trained independently on a different set of AFQMC calculations for a given lattice size. In the inset, for the $11^3$ lattice, we show the percent error as a function of the number of training points for PMMs of three different dimensions.}

\label{fig:pmm-performance}
\end{figure}

A PMM learns a subspace projection \cite{Bonilla:2022rph} of the Hamiltonian to reproduce its expectation values for a wide set of its dependent parameters \cite{Cook:2024toj}. 
To accomplish this, one creates an implicit relation between the control parameters and the calculations being emulated, allowing the model to predict results for any input values. 
The exact form of the equation can vary, and is often chosen to be similar to the original equation describing the system.
Since we are emulating the lattice Hamiltonian given by Eq.~(\ref{eq:H}), we implement a PMM in the form of, 
\begin{align} \label{eq:pmm}
    H = H_0 + \gamma H_\gamma + UH_U,
\end{align}
where $H_0$ is a diagonal matrix, $H_\gamma$ and $H_U$ are symmetric matrices, and $\gamma$ and $U$ are the control parameters the original Hamiltonian is dependent on.
The elements of these matrices are composed of trainable parameters found by fitting the lowest real eigenvalue of the resulting matrix to the data being emulated.
The dimensionality of the matrices can be chosen as a hyperparameter of the model. A higher dimensioned PMM means that there are more parameters to train which can lead to higher accuracy at a higher computational cost. 
Following Refs.~\cite{Armstrong:2025tza, Curry:2025pna}, we tested PMM dimensions from 3 to 5, as shown for one lattice size in the inset to Fig.~\ref{fig:pmm-performance}. We found that a 4-dimensional PMM gives an acceptable level of accuracy while keeping the computational cost reasonable.

From a set of 100 AFQMC calculations, the first training point is chosen to be that of the highest energy \cite{Armstrong:2025tza}. 
The remaining points are randomly divided into a validation set and a test set. 
The fitting routine is constructed iteratively by choosing the initial guess to be the result of the previous fit. 
After each optimization, the next training point candidate is selected from the validation set to be the one with the maximum percent error between the PMM output and the given data. 
It is accepted if the average percent error across the validation set decreases; otherwise it is rejected and a new candidate is selected using the same criteria until this condition is met. 
Using this routine with the AFQMC data, we are able to tune our PMMs such that they exhibit a $\lesssim 1\%$ error when compared to the test set withheld from the model shown in Fig. \ref{fig:pmm-performance}. 

\begin{figure}[t]
\centering
    \includegraphics[width=0.49\textwidth]{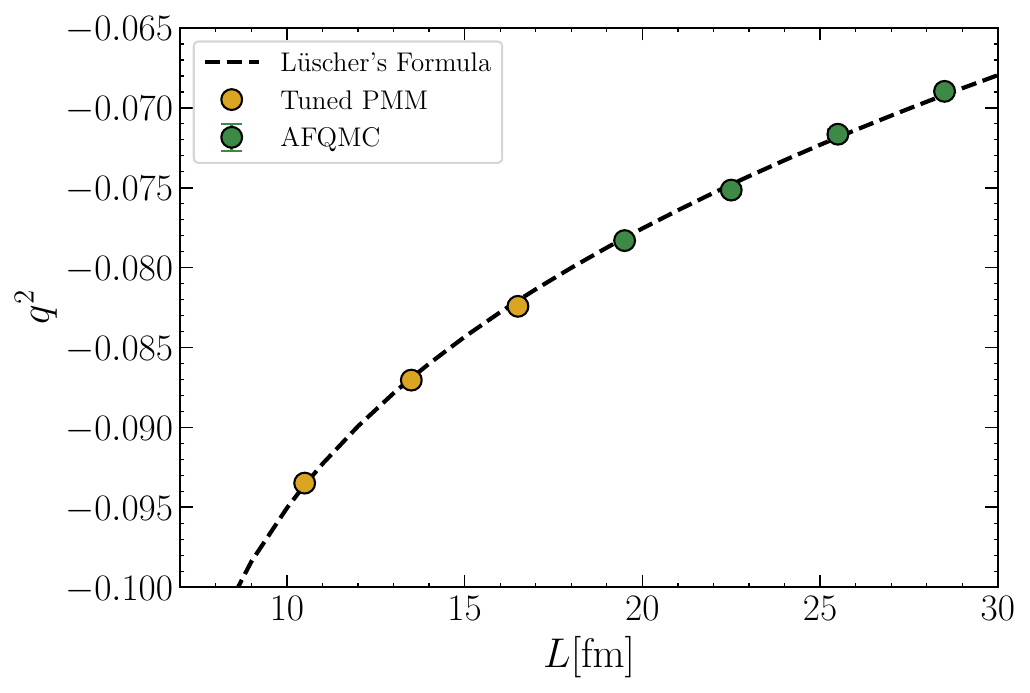}
    \caption
{
	Schematic description of our new approach to tuning the lattice Hamiltonian parameters (for $\alpha=1.5\ \text{fm}$) to reproduce known scattering length and effective range parameters. PMMs that have been trained on successively larger lattice sizes emulate AFQMC results, $q^2 = (Lp/2\pi)^2 = EL^2/4m\pi^2$, for any choice of $\gamma$ and $U$ until they reproduce the exact L\"{u}scher's formula results. These parameters are then used for AFQMC many-body calculations of larger lattices, showing excellent agreement with the exact result. 
}
\label{fig:luscher}
\end{figure}

After tuning our emulators for three different lattice sizes, we are able to use them to rapidly evaluate the AFQMC estimate of the two-body ground-state energy for different values of $\gamma$ and $U$ across the parameter space. We can then tune these parameters to reproduce the energies coming from L\"{u}scher's formula as discussed above. To test this procedure we have tuned our interaction to reproduce the neutron-neutron s-wave scattering length and effective range, see Fig. \ref{fig:luscher}, for the $7^3$, $9^3$, and $11^3$ lattices simultaneously. Then, using this interaction, carried out full AFQMC calculations of the two-neutron system at even larger lattice sizes, finding a less than 1\% disagreement with the exact L\"{u}scher's formula result, meaning the interaction is properly tuned and can be used in further many-body calculations. In the future, it would be worthwhile to explore building emulators that can also learn how the AFQMC results change as a function of lattice spacing, and to investigate extensions to L\"{u}scher's formula beyond s-wave scattering which are necessary for more complicated nuclear interactions that are required to describe finite nuclei. 

\section{From UltraCold Atoms to Nuclear Physics} \label{sec:polaron}
The Fermi polaron, defined as a single spin-down particle within a background gas of spin-up particles, has been studied experimentally in ultracold radio-frequency spectroscopy experiments \cite{Schirotzek:2009aa, Yan:2019oqn}.
We are able to study the polaron system using the many-body Hamiltonian introduced in Eq.~(\ref{eq:H}), and though the drastic spin imbalance introduces a considerable fermion sign problem, the constrained path approximation proves more than sufficient to control it. 
Due to their quasi-exact nature, Quantum Monte Carlo calculations are able to provide meaningful guidance for these and other ultracold atoms experiments \cite{Carlson:2011kv, Gezerlis:2009xp, Ku:2012utq, Kohstall:2012aa}. 

\begin{figure}[b]
\centering
    \includegraphics[width=0.49\textwidth]{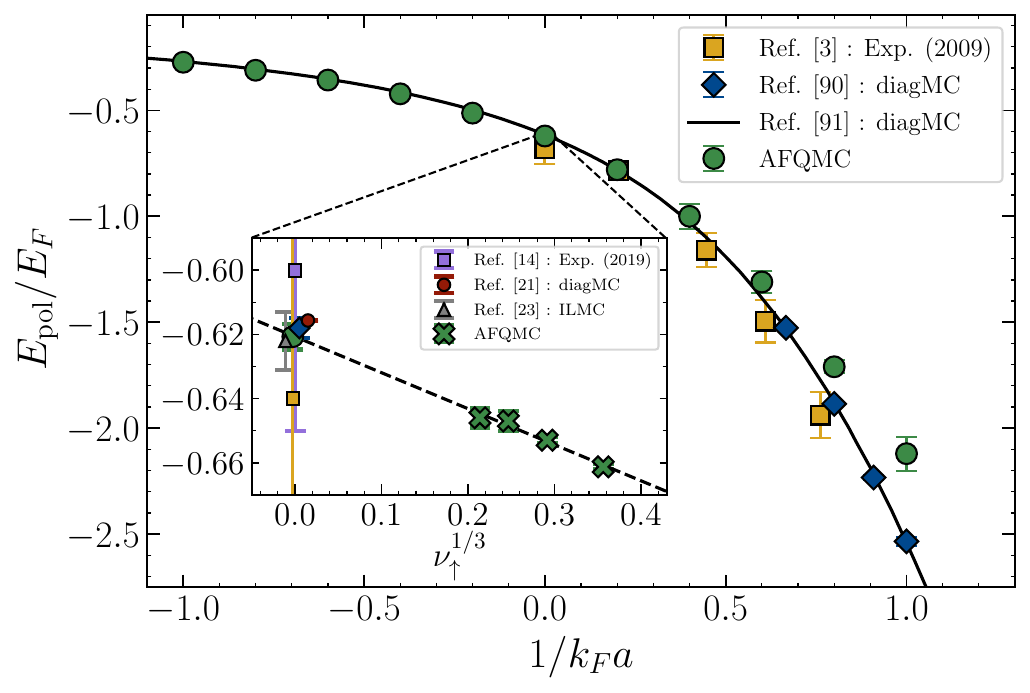}
    \caption
{
	    Energy of the Fermi polaron, in units of ${E_F = k_F^2/2m}$, as a function of $1/k_Fa$ where $a$ is the tunable scattering length of the interaction. AFQMC calculations are carried out at exactly $r_e=0$.  
 	    We compare our AFQMC calculations against previous Diagrammatic Monte Carlo calculations \cite{Prokofev:2008aa, Vlietinck:2013aa} and experimental results \cite{Schirotzek:2009aa}.
	    In the inset we also show our AFQMC calculations at unitarity, and our extrapolation to the continuum limit of zero filling factor.
	    We find excellent agreement with recent experiments \cite{Schirotzek:2009aa, Yan:2019oqn}, and other \textit{ab initio} many-body calculations such as diagrammatic Monte Carlo and impurity lattice Monte Carlo (ILMC) \cite{Prokofev:2008aa, VanHoucke:2020aa, Bour:2014bxa}. 
	    The horizontal scatter in the points at $\nu_{\uparrow}^{1/3}=0$ is purely to aid in legibility.
} 
\label{fig:coldatoms}
\end{figure}

One of the most interesting features of cold-atomic experiments, is their ability to tune the interaction between particles, allowing a study across different scattering lengths. Systems at unitarity, the limit of infinite scattering length and vanishing effective range, have been studied widely with QMC approaches and are particularly interesting as they boast a number of universal properties \cite{Carlson:2007omx, Lee:2008xsa, Blume:2007aa, Gandolfi:2011aa, Forbes:2010gt}. We are able to perform AFQMC calculations of the Fermi polaron directly at unitarity, as well as for a wide range of $k_F a$ values. 
In Fig. \ref{fig:coldatoms} we compare our AFQMC calculations of the Fermi polaron energy, $E_{\text{pol}} = E_{N_{\uparrow}+1} - E_{N_{\uparrow}}$,
against both experimental results \cite{Schirotzek:2009aa, Yan:2019oqn}, and previous theoretical calculations \cite{Prokofev:2008aa, VanHoucke:2020aa, Bour:2014bxa}. 
The AFQMC results for $E_{N_{\uparrow}+1}$ and $E_{N_{\uparrow}}$ are compared at fixed volume, 
\begin{align}
	L^3 = (\alpha M)^3 = \frac{6\pi^2N_{\uparrow}}{k_F^3},
\end{align}
and not at a fixed density. 
\begin{figure}[t] 
\centering
   \includegraphics[width=0.46\textwidth]{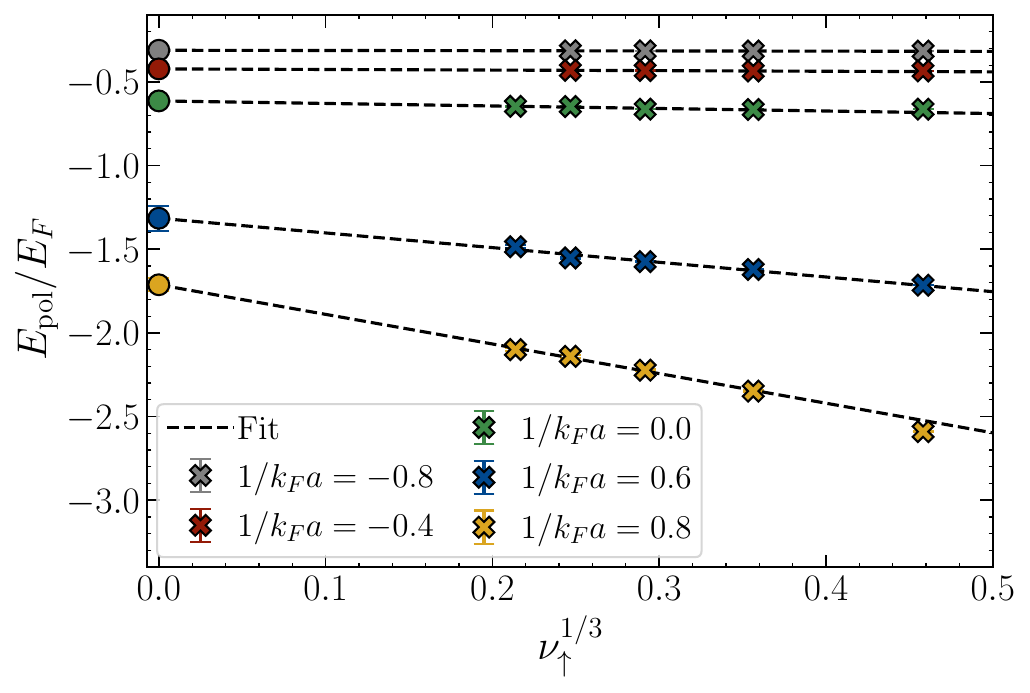}
   \caption{Extrapolation to the continuum limit of zero filling factor, $\nu_{\uparrow} = N_{\uparrow}/M^3$ for our AFQMC calculations of the Fermi polaron. 
   The number of spin-up particles is kept fixed at ${N_{\uparrow}=33}$, and we simultaneously increase the number of lattice sites while decreasing the lattice spacing to extrapolate to the continuum limit while maintaining a fixed $k_Fa$.}
\label{fig:extrap}
\end{figure}

Our result at unitarity agrees with all previous experimental and theoretical predictions. In the inset of Fig. \ref{fig:coldatoms} we explore an important feature of the AFQMC algorithm. 
Since we are taking a physical system and placing it on a discretized spatial lattice, there is some inherent error associated with this approximation that must be accounted for. 
Recent finite temperature calculations explored calculations at successively larger lattice sizes to extrapolate to the continuum limit \cite{Jensen:2019zkr}. 
We have employed similar extrapolations for all systems studied in the remainder of this work, with a selection from both sides of $k_F a=0$ shown in Fig. \ref{fig:extrap}.
In the inset to Fig. \ref{fig:coldatoms} we show explicitly our extrapolation to the continuum limit for the Fermi polaron at unitarity, which illustrates the importance of these extrapolations to provide meaningful predictions. 
The fact that AFQMC is also able to smoothly study the full crossover region in $k_F a$, without being hindered by the sign problem, is promising for future studies on polarized cold-atomic gases.

\begin{figure}[b]
\centering
    \includegraphics[width=0.49\textwidth]{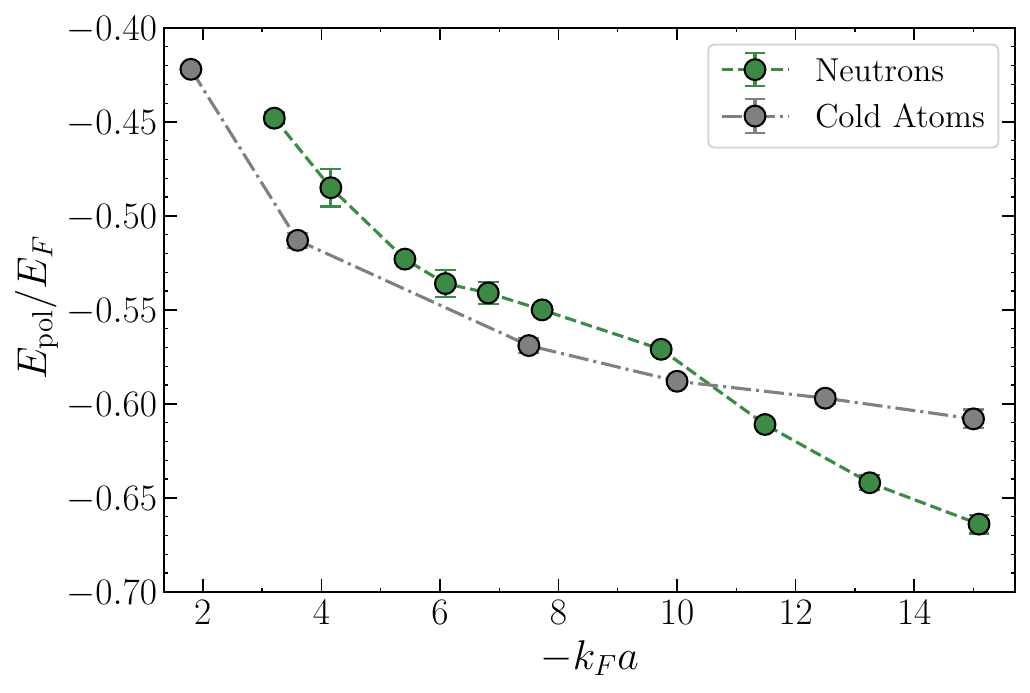}
    \caption{Energy of the Fermi and neutron polaron across a range of negative $k_F a$. The dashed lines are included only to guide the eye. All the AFQMC results are extrapolated to the continuum limit, see Fig. \ref{fig:extrap}.} 
\label{fig:neutrons_vs_coldatoms}
\end{figure}

While cold-atomic experiments are able to successfully probe the Fermi polaron, recent years have shown that the polaron is also an interesting system within the realm of nuclear physics.
The $s$-wave scattering length for the neutron-neutron interaction, $a \approx -18.5\ \text{fm}$, is often characterized as unusually large, 
which leads to a natural comparison with cold-atomic experiments at unitarity. 
Though both the unitary Fermi gas and neutron matter have large negative scattering lengths, they differ dramatically in their effective ranges. 
Our cold atoms calculations in Fig.~\ref{fig:neutrons_vs_coldatoms} are carried out at unitarity ($r_e=0$) the neutron-neutron s-wave interaction has a reasonable large finite effective range of roughly $2.7\ \text{fm}$.
We have performed calculations for the neutron polaron, a spin down neutron immersed in a background gas of noninteracting spin up neutrons, over a range of $k_F$. To meaningfully compare these calculations with our results for cold atoms, we plot both the Fermi and neutron polaron in Fig. \ref{fig:neutrons_vs_coldatoms} as a function of $-k_F a$, which has previously been used to contrast these distinct physical systems \cite{Gezerlis:2007fs, Gezerlis:2009iw}.
We find that in the regime of small $k_F a$, low density neutrons and weakly interacting cold atoms, the cold atomic polaron exhibits a much larger magnitude polaron energy. However, as the cold atomic interaction is increased and the neutron polaron system is moved to higher densities, we eventually encountered a crossover region where after a certain point it is the neutron polaron whose polaron energy has the larger magnitude.

\begin{figure}[t]
\centering
    \includegraphics[width=0.49\textwidth]{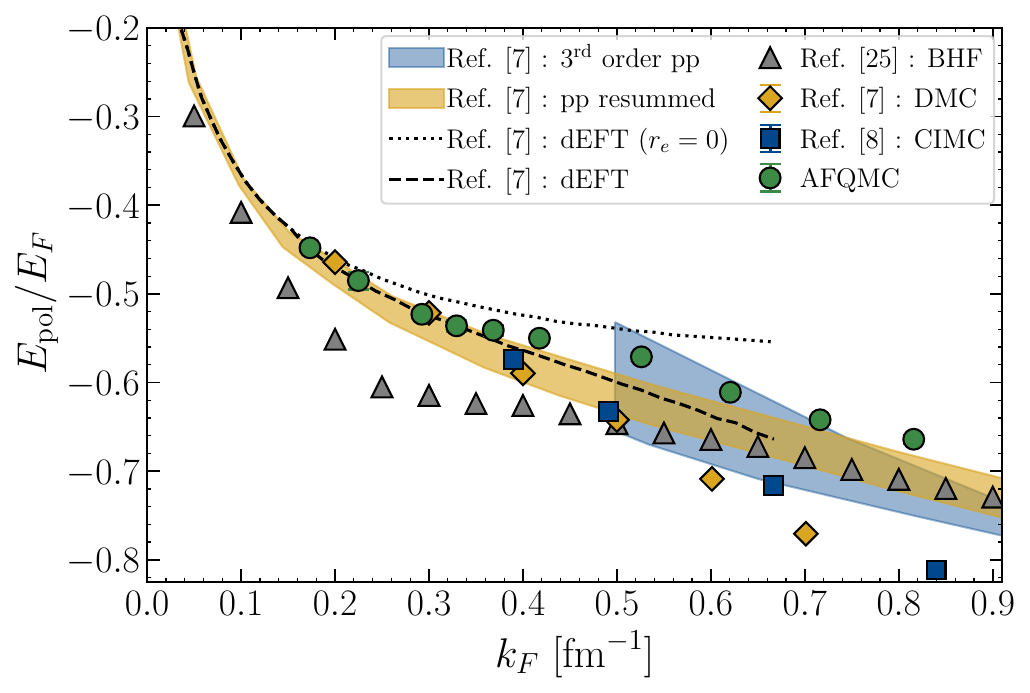}
    \caption{Energy of the neutron polaron in units of the Fermi energy as a function of the Fermi wave number. 
We compare our AFQMC calculations against previous Diffusion Monte Carlo and chiral effective field theory calculations from Ref.~\cite{Forbes:2013ava}, configuration interaction Monte Carlo calculations from Ref.~\cite{Roggero:2014cma},  as well as Brueckner-Hartree-Fock calculations using two different nuclear interactions from Ref.~\cite{Vidana:2021tiw}.}
\label{fig:neutrons}
\end{figure}

We are also able to place our neutron polaron calculations within the context of previous theoretical investigations. 
As shown in Fig.~\ref{fig:neutrons}, we are able to compute the energy of the neutron polaron across a wide range of densities, 
which allows us to compare our calculations with Ref.~\cite{Forbes:2013ava}, which performed both \textit{ab initio} Diffusion Monte Carlo (DMC) calculations as well as effective field theory calculations,
Ref. \cite{Roggero:2014cma} which performed similar calculations using the Fock space configuration interaction Monte Carlo (CIMC), and 
Ref. \cite{Vidana:2021tiw} which was able to explore a much larger region of $k_F$ with a phenomenological Bruecker-Hartree-Fock (BHF) approach. 

In the low-density region we find excellent agreement with the previous DMC calculations, and are able to provide predictions at even smaller $k_F$ values than were possible using the coordinate space approach. To do this, we needed to employ lattices with up to $23^3$ lattice sites, and particle numbers from $N_{\uparrow}=7$ up to $N_{\uparrow}=57$, which involved considerable computational resources. 

We interestingly find a divergence from both the \textit{ab initio} and phenomenological results beyond densities of $k_F \approx 0.4\ \text{fm}^{-1}$, however our results agree reasonably well with the bounds provided by Ref.~\cite{Forbes:2013ava}'s chiral EFT bands.
This disagreement is likely due to differences in the interactions used in this work and those in Refs.~\cite{Forbes:2013ava, Roggero:2014cma}, in particular the unknown shape parameter from Eq.~(\ref{eq:ere}). 
Given our free projection calculation in Fig.~\ref{fig:constrained}, a conservative estimate of this effect would be on the order of a few percent, however a detailed study also including DMC fixed- and released-node calculations would be required to fully explore this effect. Based on our results, it is reasonable that in the future one could potentially use the polaron to investigate how DMC and AFQMC differ in their approach. 

The polaron calculations in this section, in both the regimes of cold-atomic and nuclear physics, are evidence of the strength of the AFQMC method. Even though the polaron is a system with a strong fermion sign problem due to the population imbalance, the constrained path AFQMC method is still able to provide meaningful and accurate predictions. In addition, the fact that AFQMC can study these diverse systems using a single Hamiltonian is a good indication that it can be applied to more complicated cold-atomic and nuclear systems.

\section{CONCLUSION}
We have performed lattice AFQMC calculations of the polaron in the context of cold-atomic and nuclear physics. To do this we have employed extrapolations to the continuum limit and begun investigations into the use of emulators in pursuit of the tuning of our lattice interactions.
We find very good agreement with previous experimental and theoretical studies of the polaron.
Though the neutron and proton polarons have been used in the past to provide constraints for energy density functionals, 
it would be an interesting question to investigate whether they could be used in a similar fashion to constrain other phenomenological calculations or even interactions such as those derived from chiral effective field theory. 
In addition, there is a natural comparison to be made between the polaron and physics that emerges in the study of nuclei, such as clustering or halo nuclei.
Future work could explore how our AFQMC calculations of the polaron could be used as a starting point for investigations into these more exotic nuclear phenomena.

\section*{Acknowledgements}
RC thanks I. Tews and H. Tajima for insightful discussions, and the authors thank I. Vida\~na, N. Prokof'ev, B. Svistunov, K. Van Houcke,  and A. Roggero for sharing the results of their previous work. 
This work was supported by the Natural Sciences and Engineering Research Council (NSERC) of Canada and the Canada Foundation for Innovation (CFI). Computational resources have been provided by Compute Ontario through the Digital Research Alliance of Canada, and by the National Energy Research Scientific Computing Center (NERSC), which is supported by the U.S. Department of Energy, Office of Science, under contract No. DE-AC02-05CH11231. 

\bibliography{biblio}
\end{document}